\documentclass[conference]{IEEEtran}

\IEEEoverridecommandlockouts

\usepackage{tikz}

\newcommand\copyrighttext{%
  \footnotesize \textcopyright 2021 IEEE. Personal use of this material is permitted. Permission from IEEE must be obtained for all other uses, in any current or future media, including reprinting/republishing this material for advertising or promotional purposes, creating new collective works, for resale or redistribution to servers or lists, or reuse of any copyrighted component of this work in other works.}
\newcommand\copyrightnotice{%
\begin{tikzpicture}[remember picture,overlay]
\node[anchor=south,yshift=10pt] at (current page.south) {{\parbox{\dimexpr\textwidth-\fboxsep-\fboxrule\relax}{\copyrighttext}}};
\end{tikzpicture}%
}

\newcommand\acceptedtext{%
  \footnotesize Accepted at The 30th International Conference on Computer Communications and Networks (ICCCN 2021)}
\newcommand\acceptednotice{%
\begin{tikzpicture}[remember picture,overlay]
\node[anchor=north,yshift=-30pt] at (current page.north) {\acceptedtext};
\end{tikzpicture}%
}

\usepackage{cite}
\usepackage{amsmath,amssymb,amsfonts}
\usepackage{algorithmic}
\usepackage{graphicx}
\usepackage{textcomp}
\usepackage{xcolor}
\def\BibTeX{{\rm B\kern-.05em{\sc i\kern-.025em b}\kern-.08em
    T\kern-.1667em\lower.7ex\hbox{E}\kern-.125emX}}

\graphicspath{{figures/}}

\ifCLASSOPTIONcompsoc
	\usepackage[caption=false,font=normalsize,labelfont=sf,textfont=sf]{subfig}
\else
	\usepackage[caption=false,font=footnotesize]{subfig}
\fi

\begin{document}

\title{On Data-centric Forwarding in Mobile Ad-hoc Networks: Baseline Design and Simulation Analysis
}

\author{\IEEEauthorblockN{Md Ashiqur Rahman}
\IEEEauthorblockA{
	\textit{The University of Arizona}\\
	marahman@cs.arizona.edu}
\and
\IEEEauthorblockN{Beichuan Zhang}
\IEEEauthorblockA{
	\textit{The University of Arizona}\\
	bzhang@cs.arizona.edu}
}

\maketitle

\copyrightnotice
\acceptednotice

\vspace{-10pt}
\begin{abstract}
	IP networking deals with end-to-end communication where the network layer routing protocols maintain the reachability from one address to another. However, challenging environments, such as mobile ad-hoc networks or MANETs, lead to frequent path failures and changes between the sender and receiver, incurring higher packet loss. The obligatory route setup and maintenance of a device-to-device stable path in MANETs incur significant \textit{data retrieval delay and transmission overhead}. Such overhead exaggerates the packet loss manifold.
  
  Named Data Networking (NDN) can avoid such delays and overhead and significantly improve the overall network performance. It does so with direct application-controlled named-data retrieval from any node in a network instead of reaching a specific IP address with protocol message exchange. However, existing works lack any explicit or systematic analysis to justify such claims. Our work analyzes the core NDN and IP architectures in a MANET at a baseline level. The extensive simulations show that NDN, when applied correctly, yields much lower data retrieval latency than IP and can lower the network transmission overhead in most cases. As a result, NDN's \textit{stateful forwarder} can significantly increase the retrieval rate, offering a better trade-off at the network layer. Such performance comes from its caching, built-in multicast, and request aggregation without requiring an IP-like separate routing control plane.
\end{abstract}


\section{Introduction}
\label{sec:introduction}

The use of broadcast-based self-learning in an IP-based network is a reasonably common technique, broadcasting the first packet and waiting for a response. The returning response packet allows the network to learn a path between a sender-receiver pair for future packet exchange. However, unlike in a wired network, wireless mobile ad-hoc networks (MANET) present significant communication challenges. Not only do they have issues like wireless channel contention and interference, but they also suffer from frequent link breakage due to mobility. As a result, \cite{Perkins:1999:AODV} proposed the AODV routing protocol for MANETs, designed on top of the broadcast-based self-learning philosophy. It was also a reactive routing protocol that minimizes the frequent routing message exchange in proactive routing protocols (e.g., OLSR~\cite{JacquetOLSR2001}). Although proactive protocols exchange these messages to compute and maintain forwarding paths, they cause significant collision and contention in the wireless channel.

AODV reduces the proactive routing overhead by only sending control messages when there is traffic to transmit. However, path discovery and maintenance still lead to added latency \textit{before} actual data exchange can occur. Thus, latency also plays a crucial role in the overall MANET performance because wireless links can break between path discovery and data exchange. Higher latency and transmission overhead from discovery and maintenance lead to higher packet loss. While there has been lots of research in this area, most of the proposed solutions are still built on top of the traditional IP architecture, which requires the discovery and maintenance of a working path between the sending and receiving hosts. Maintaining such working paths is inherently challenging in a MANET environment.

Named Data Networking (NDN)~\cite{ndn-ccr} is an emerging Internet architecture that moves away from IP's address-based point-to-point communication model to a data-centric communication model. NDN packets do not carry any address; instead, they carry the name of the content that consumer nodes want to retrieve. A consumer sends an Interest (i.e., a request packet) to the network, and the network returns a Data (i.e., a response packet) whose name matches what was requested by the Interest. Data can come from any node in the network, including data producers and caches. NDN's broadcast-based self-learning can be a better fit for MANET than IP because (1) data discovery and link breakage detection can be achieved using the Interest/Data exchange without extra control messages to reduce collision, contention, and most importantly, retrieval latency; (2) since data can be cached and retrieved from any node, the Interest does not have to reach the original producer; thus, the impact of link/path breakage can be significantly mitigated; (3) multicast is achieved in NDN without any extra mechanism, which is a tremendous advantage since multicast in IP MANET is still a challenging problem.

The self-learning forwarding by \cite{Junxiao:SelfLearning:2017} is also a broadcast-based discovery technique employing NDN as the underlying architecture. Their work shows how NDN can be deployed in a wired local network and suggests a promising MANET design. There have also been some research proposing NDN-based MANET routing and forwarding protocols \cite{etefia2012named, azgin2014mobility, teixeira2016can, khelifi2019named}. However, the impact of the architecture on MANET forwarding performance has not been systematically investigated nor fully understood. In this paper, we conducted comprehensive simulations to compare the packet delivery performance of NDN and IP in a MANET at the network layer and bridge the gap between \textit{promises} and \textit{simulated results} for NDN in MANET. We use AODV~\cite{Perkins:1999:AODV} in IP and our proposed Data-centric Ad-hoc Forwarding (DAF) in NDN as the network-layer protocol for routing/forwarding. These two protocols use similar mechanisms to discover paths or contents, and we do not apply any optimization techniques beyond the basic protocol. Therefore our results provide a baseline comparison between NDN and IP, reveal their architectural differences, and point to further optimizations.

Our simulations show DAF outperforming AODV in this aspect, especially under mobility. The main reason is that DAF does away with routing protocol by relying on Interest/Data exchange to discover data and detect failures. It significantly reduces the latency and, in most cases, the transmission overhead as well, which are two of the most challenging factors in an IP-based MANET routing. As a result, DAF vastly improves the data retrieval success rate by properly utilizing NDN as its underlying architecture. The benefits of NDN are even more prominent for many-to-one and many-to-many communication due to its built-in caching and multicast.

The evaluation of NDN and IP's performance in the MANET environment illustrates their architectural differences. It demonstrates that NDN's data-centric design is a better fit for MANET, where maintaining paths between two moving nodes is inherently challenging and sometimes unnecessary. With a new underlying architecture, many potential techniques can be considered to optimize NDN-based MANET performance in the future.

\section{Related Works}
\label{sec:related-works}

Routing in MANETs has been one of the most challenging and well-studied topics in computer networks. \cite{boukerche2011routing} presents a comprehensive classification of ad hoc routing protocols, e.g., reactive, proactive, hybrid, geolocation aware, to name a few. DSDV~\cite{Perkins:1994:HDD:190809.190336}, OLSR~\cite{JacquetOLSR2001}, GSR~\cite{Chen:GSR:1998} are some examples from the proactive class. They exchange periodic routing updates to calculate address-based routes quickly. However, it incurs significant bandwidth usage, collision, and contention and is unsuitable under moderate to high mobility.

Reactive routing protocols such as DSR~\cite{Johnson1996} and AODV~\cite{Perkins:1999:AODV} reduce the proactive routing overhead by exchanging routing information only before data transfer. However, they suffer from route setup and maintenance delay, which gets worse with increasing mobility. Hybrid approaches like ZRP~\cite{haas2001zone}, DST~\cite{radhakrishnan1999dst} either suffer from high computational overhead or single point of failure, respectively. GPSR~\cite{Karp:2000:GGP:345910.345953} has quick route maintenance but high computational overhead from frequent neighbor updates and requires geolocation information. REEF~\cite{conti2006reliable} selects the next hop in a probabilistic manner but still needs to reach a specific destination address.

Another major challenge in IP networking is multicast support to reduce network overhead in group communication. Deploying multicast over IP in a regular wired network is already tricky enough~\cite{diot2000McastDeployment}. The multicast tree needs to keep track of the endpoints or IP addresses and maintain a path towards them. Re-adjusting the tree when members join or leave introduces additional overhead, which is more frequent in MANETs, leading to further complexities~\cite{ZianeAMDR2007, Bae:GeoTora:2000, royer1999multicast}. \cite{de2003multicastManet} dissects multicast over MANETs, pointing towards specialized protocols for specific environments making it even harder to find common ground. All these protocols try to minimize the routing overhead, but with the IP's address-based end-to-end path setup and maintenance, they all require substantial modifications and end up with intricate designs. Without a specialized multicast routing, IP requires dedicated flow per sender-receiver pair for the same content, leading to high transmission overhead and high collision and contention.

On the contrary, NDN's \textit{data first} scheme supports multicast by design and enables caching to reduce network load and increase retrievability. \cite{teixeira2016can} shows a comparative analysis between NDN and OLSR in MANETs but lacks any forwarding improvement and motivation of why and how NDN can be a better solution than IP. LFBL~\cite{Meisel:2010:LFBL} uses wireless devices' broadcast nature as an advantage for their producer-independent data retrieval. However, it considers specific consumer and producer id-based communication and does not include caching and aggregation, two fundamental components of NDN's design. Implementation of LFBL~\cite{amadeo2015forwarding} also shows a very high hop-by-hop forwarding delay, terrible for mobile networks. Performing pure broadcast in a wireless network also incurs the potential overhead of high CPU processing. Self-learning in \cite{Junxiao:SelfLearning:2017} promises possible transmission event reduction. However, it yields very high-Interest broadcast and packet loss under mobility as it only performs Interest broadcast from a consumer upon Interest timeout. Thus many unicast Interests do not reach the next hop from mobility, and the consumer ends up sending a high number of broadcast Interests for discovery.

Forwarding or retransmission decision based on RTT measurement is present in existing NDN literature. The hyperbolic routing \cite{lehman2016experimental} uses an adaptive smoothed RTT-based forwarding (ASF) as a sub-module to minimize the sub-optimal path selection. However, overhead from periodically probing individual interfaces and high convergence time on link failure is unsuitable for mobile networks. DAF avoids such complexities and employs a quicker reaction mechanism for a better baseline analysis of NDN and IP architectures at the network level.
\section{System Model}
\label{sec:system-model}

\begin{figure}[!t]
  \centering
  \includegraphics[width=0.98\columnwidth]{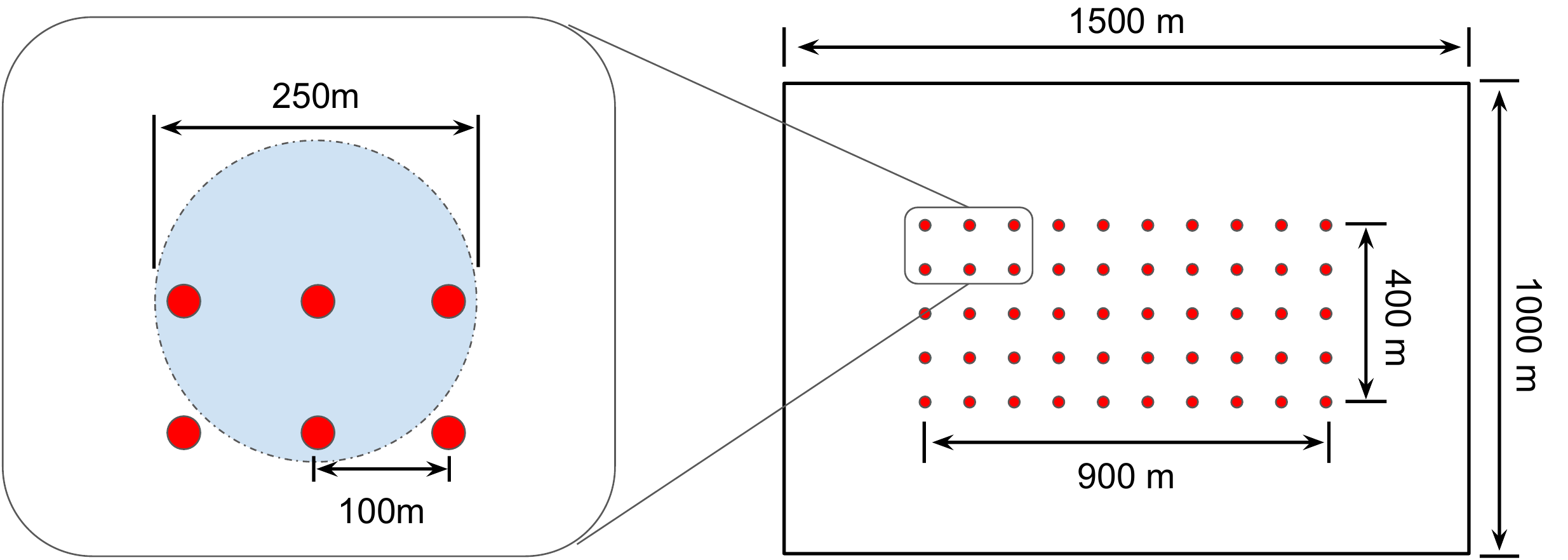}
  \caption{Initial system setup for 50 nodes when all nodes are stationary. The right side shows the grid topology setup, and the left side shows inter-node distance and transmission range (colored circular area).}
  \label{fig:system-example}
\end{figure}

Our MANET's architectural analysis employs the general system model following the random geometric graphs~\cite{dall2002random}. The only difference is that at the beginning of a simulation, instead of placing nodes at random points in the 2-dimensional plane, we place the stationary nodes randomly in a grid topology with either 10x5 or 10x10 setup for 50 and 100 nodes network, respectively. We use this setup to analyze the stationary wireless network behavior when a client or consumer can always reach a server or a producer. Each node has a fixed transmission range of 250 meters in diameter. When stationary (speed=0 m/s), we place the nodes 100 meters apart in X and Y coordinates to communicate only with their immediate neighbors at the top, bottom, left, and right (maximum four). For 100 nodes, we placed them in a 900m x 900m 2D grid (10x10) within an area of 1500m x 1500m. For 50 nodes, we placed them in a 900m x 400m grid (10x5) within an area of 1500m x 1000m. One example setup of the system model for 50 nodes topology is in Fig.~\ref{fig:system-example}. 

In simulations with mobility, nodes start to move around at time $t>0$, and their speed is also set to $>0$ m/s. We use the random walk 2d mobility model, with 5 seconds of mobility, and without pause before setting the next random point. This basic mobility helps us to establish a baseline analysis. Scenario specific mobility model (e.g. manhattan or circular mobility) may impact the forwarding performance which we leave for future work. We also consider nodes' mobility in various simulations to range between 0 m/s to 8 m/s. Each node has only one wireless interface communicating over a single channel of IEEE 802.11b at 11Mbps. The system consists of 10 consumers or clients and either 10 or 2 producers or servers with the remaining nodes as packet forwarders. Thus we conduct experiments with 1-to-1, many-to-many, and many-to-1 communication. In IP, they are trivial as one client IP address can communicate with only one server address. However, in NDN, any Data node (producer or cache) can serve a consumer following its address and location-agnostic design.
\section{Design Backgrounds}
\label{sec:motivation-background}

While a network architecture provides the backbone model for communication, a routing protocol or forwarding strategy utilizes the backbone to make packet forwarding decisions. Thus, we go over the background and understand the general idea of AODV routing, the general NDN architecture, and some existing forwarding designs. 

\textbf{IP-AODV Routing}: Like most IP routing for MANETs, AODV~\cite{Perkins:1999:AODV} also exchanges protocol messages to ensure an end-to-end path or route between source-destination addresses before sending data packets. These protocol messages ensure a \textit{loop-free, shortest-path} packet forwarding between the source and destination addresses as the IP forwarding plane cannot detect any loop by itself. AODV follows a broadcast-based discovery mechanism to learn the shortest path. When a source node's application wants to initiate data exchange, AODV buffers the packet and broadcasts route-request (RREQ) to the network. On an RREQ, the destination address, or any network with an existing route to the destination, send back a route-reply (RREP) as unicast. The intermediate nodes and eventually, the source node, learn the shortest path to the destination IP address from multiple RREP coming from possible multiple neighboring nodes. AODV then forwards the buffered packet over the learned path.

AODV ensures end-to-end path availability by exchanging periodic \verb|HELLO| messages between nodes on an active route. In the case of a path-break from \verb|HELLO| packet loss, network nodes issue route-error (RRER) packets back to the source node to trigger RREQ broadcast. A node never forwards data in the absence of a valid path. Thus, route setup and maintenance induce delay before and during data exchange.

Given that AODV offers a straightforward routing over the IP model without any further added complexities such as multicast or using geolocation information, we consider it as a suitable baseline. 

\begin{figure}[!t]
  \centering
  \includegraphics[width=0.98\columnwidth]{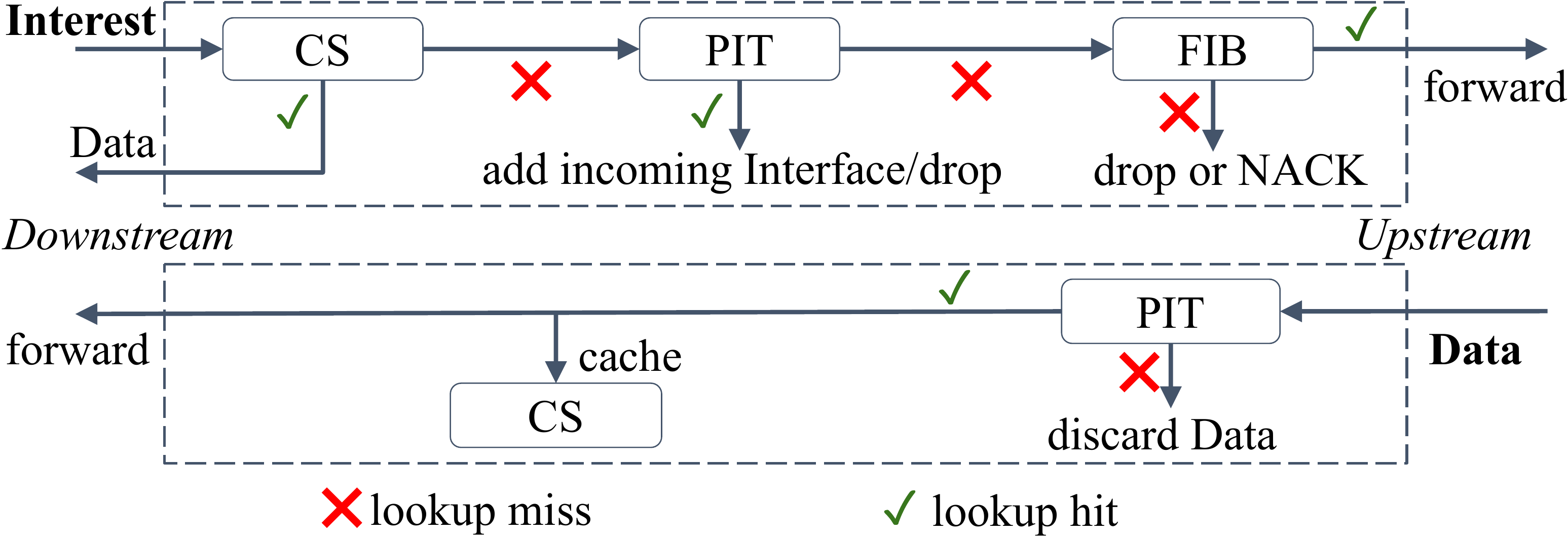}
  \caption{Stateful forwarding pipeline in an NDN node.}
  \label{fig:ndn-forwarding}
\end{figure}

\textbf{Named Data Networking}: Fig.~\ref{fig:ndn-forwarding} shows the general architecture of an NDN node. A consumer sends an Interest with a name, e.g., \verb|/news/image.png|. Upon receiving, a node looks at its Content Store (CS) and replies with data on a name match. Otherwise, the node passes the Interest to its Pending Interest Table (PIT), and if there is a previous entry, it suppresses the Interest with necessary lifetime updates (e.g., 2s lifetime). Otherwise, the node creates a new PIT entry with the incoming interface and passes the Interest to its Forwarding Information Base (FIB). There, it performs a longest common prefix (LCP) match to existing FIB prefix entries. On a FIB match, e.g., \verb|/news|, a node forwards the Interest through the associated interface(s) to the next-hop node(s). Otherwise, it either drops the Interest or sends a \verb|NACK| (negative acknowledgment) back to the downstream. A \verb|NACK| informs the downstream node that a next-hop in the upstream is not available and that it may try to forward the Interest to the next available FIB entry.

PIT also helps with loop-free Interest forwarding from a packet's \verb|NONCE|. When a node receives data, it first checks the PIT, and on a match, passes the data to the CS; otherwise drops. After updating CS, the node forwards data to the associated downstream(s) and clears the PIT entry.


\textbf{NDN flooding}: An NDN flooding strategy follows the above design but broadcasts all Interests regardless of a FIB hit or miss. Thus, flooding can benefit from multiple paths between a consumer and data node(s), increasing data retrieval possibility. However, a data broadcast may satisfy multiple nodes' PITs resulting in a broadcast storm. It leads to significantly high collision and channel contention in the network. There is also no \verb|NACK| involved as every packet transmission is broadcast. Flooding data in IP without a routing protocol is impossible as it lacks built-in loop detection.


\textbf{NDN self-learning~\cite{Junxiao:SelfLearning:2017}}: It can minimize the broadcast effect of flooding and is very close to IP-AODV in terms of learning and then performing unicast. A consumer broadcasts (discovery) Interests like RREQ until it \textit{learns} a next-hop node towards data. Data travels as unicast towards the consumer like RREP in a hop-by-hop manner populating downstream nodes' FIBs with prefix announcement. A node then unicasts Interests to a next-hop on the shortest path. In a wired network, self-learning can send \verb|NACK| back to a consumer, similar to RRER in AODV. It is one of the reasons we consider AODV as an IP-baseline in our network layer evaluation. 

However, sending \verb|NACK| in a MANET by self-learning is expensive for three reasons, a) a node does not know which upstream generated it, b) \verb|NACK| can also get lost from mobility, causing Interest timeout, and c) it can add transmission overhead similar to protocol message exchange. Moreover, unlike AODV's in-network RREQ broadcast, self-learning drops unicast Interest on a FIB mismatch. It may also unicast an Interest to a stale FIB entry. As only the consumer makes discovery decisions, it suffers from many application-level timeouts, and the reaction to mobility gets delayed, which is undesirable.

To mitigate the challenges and limitations of flooding and self-learning, we propose the DAF strategy as an NDN-baseline to reduce excessive network transmission and offer better mobility support. In our simulations, DAF outperforms AODV with a better application success rate by avoiding separate routing protocol and thus lowering latency and, in most cases, the overall network transmission events. Such improvements in a MANET come from NDN's architectural advantages over IP, using data names for communication instead of IP addresses.

\section{The Data-centric Ad-hoc Forwarding (DAF)}
\label{sec:daf-strategy}

We present a data-centric ad-hoc forwarding (DAF) strategy that finds a balance between flooding and self-learning to establish an NDN baseline in MANETs. It is similar to the well-known broadcast-based discovery. Moreover, DAF only \textit{manipulates} the NDN architecture innovatively to make it suitable for MANETs, just like AODV does the same in IP-based MANETs. Both AODV and DAF try to learn a next-hop over the shortest path to an IP-address and data name, respectively. However, AODV learns a route before sending or requesting data, whereas DAF uses NDN's Interest packet to broadcast a request to the network. DAF learns next-hop(s) towards the data node(s) like the self-learning strategy. However, instead of a consumer timeout-based broadcast, DAF relies on an in-network \textit{feedback-reaction} mechanism and opportunistically reacts to possible failures in a decentralized manner. 

\textbf{\textit{Positive feedback}}: Receiving a Data packet denotes positive feedback. It means that a neighbor can serve a name prefix, thus send future Interest as a unicast to that next-hop. Positive feedback creates a new FIB entry or updates an existing one. 

\textbf{\textit{Negative feedback}}: When a learned next-hop fails to reply with data within a specific time, it acts as negative or implicit feedback. It indicates a collision, contention, or link failure at any hop between the current and the data node. Negative feedback removes the expected next-hop from FIB and helps to avoid protocol messages to minimize packet collisions.

\subsection{In-network RTT measurement and FIB-next-hop lifetime}
Given that our DAF strategy does not use protocol messages for neighborhood maintenance, we need a way to detect path failure between consumer and data, potentially resulting from mobility. It allows any node to remove a next-hop neighbor towards the data from its FIB entry and fall-back into discovery mode for Interest forwarding when no other neighbor is present. Thus we set a timeout value or lifetime for each FIB next-hop entry. To calculate the lifetime of a FIB next-hop entry, we propose a decentralized in-network Round-Trip Time (RTT) measurement-based technique that alleviates NDN's concept of using the name as content identifier. It requires minimal changes to the existing NDN forwarding tables for storing the associated timestamps with negligible processing overhead. Any node in the network can perform calculations independent of the actual consumer, network diameter, and data nodes, which are some of NDN's data-centric design's primary goals. As this is a software-based technique, it applies to any wireless communication technology.
A node calculates the RTT of the $i$-th data satisfying the $i$-th Interest locally as,
\begin{equation}
  RTT_i = T(D_i) - T(R_i).
  \label{eq:rtt}
\end{equation}
Here, $T(R_i)$ and $T(D_i)$ are the Interest transmission (request) and satisfaction (on Data) timestamps, respectively. Upon receiving $D_i$ for $R_i$, a node initializes or updates the smoothed RTT ($SRTT(nh_p)$) and RTT variation ($RTTV(nh_p)$) of the FIB-next-hop entry, respectively, similar to \cite{TCP-RTx},
\begin{equation}
  SRTT(nh_p)= 
\begin{cases}
  RTT_i,& \text{if } \text{new entry}\\
  \begin{aligned}
    & \alpha.SRTT(nh_p) + \\ & (1-\alpha).RTT_i,
  \end{aligned}
  & \text{otherwise.}
\end{cases}
\label{eq:srtt}
\end{equation}

\begin{equation}
  RTTV(nh_p)= 
\begin{cases}
  \frac{RTT_i}{2},& \text{if } \text{new entry}\\
  \begin{aligned}
    & \beta.RTTV(nh_p) + (1-\beta). \\ & |RTT_i - SRTT(nh_p)|,
  \end{aligned}
  & \text{otherwise.}
\end{cases}
\label{eq:rttv}
\end{equation}

Here, $nh_p$ is the FIB-next-hop which replied with data containing prefix \verb|/p|, indicating \verb|/p| is reachable through that $nh$. We also set $\alpha=7/8$ and $\beta=3/4$. Performing such in-network RTT measurement in IP is a much complex task and requires substantial modifications. It is because IP routers are only aware of addresses and not the actual content names.

When a node unicasts an Interest (e.g., \verb|/p/a|) to a FIB-next-hop $nh_p$, it starts a timer,
\begin{equation}
 Timeout(nh_p) = SRTT(nh_p) + 4 \times RTTV(nh_p)
 \label{eq:timeout}
\end{equation}
Here, $Timeout(nh_p)$ is the lifetime of $nh_p$ in the FIB. The added variance in Equation~\ref{eq:timeout} helps to minimize the RTT fluctuations. This fluctuation can occur very often in an NDN-based MANET as any node in the network with data can reply. 
If the unicast Interest forwarder does not get data back through that $nh_p$ within $Timeout(nh_p)$, it treats it as negative feedback and removes the related FIB-next-hop entry.

\subsection{Packet Forwarding}
\label{sec:daf-forwarding}
\begin{figure}[!t]
  \centering
  \includegraphics[width=0.98\columnwidth]{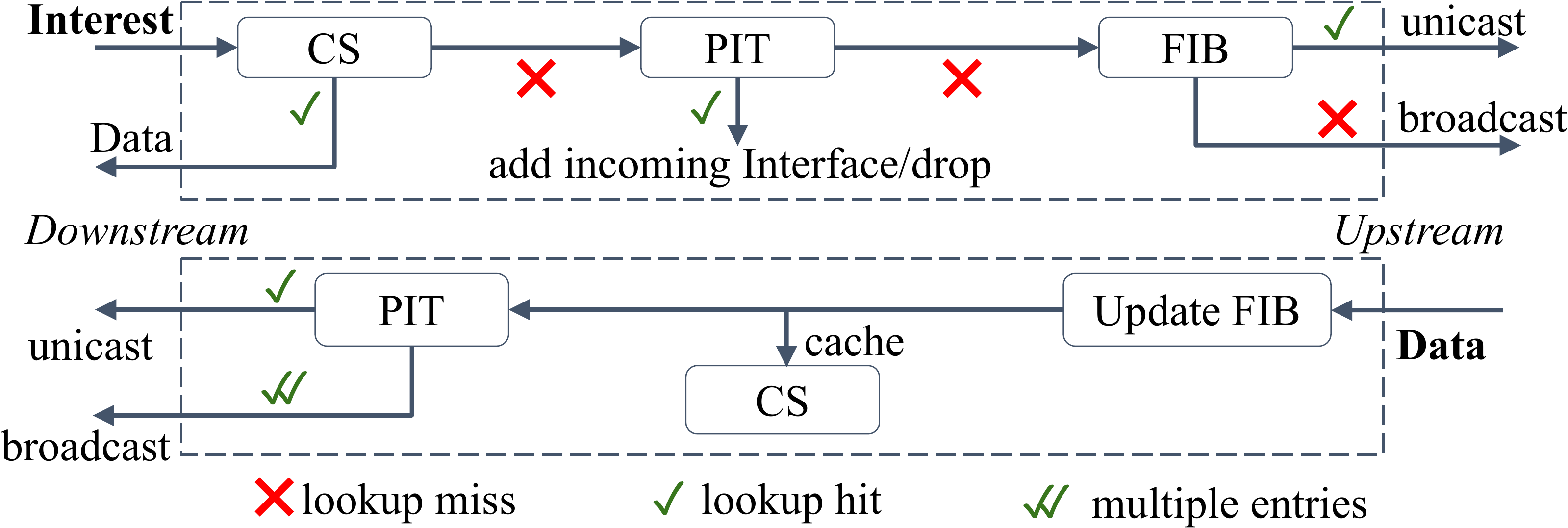}
  \caption{Forwarding pipeline of the DAF strategy for MANETs.}
  \label{fig:daf-forwarding}
\end{figure}

We show our DAF forwarding pipeline in Fig.~\ref{fig:daf-forwarding} and discuss the modifications made over Fig.~\ref{fig:ndn-forwarding}.

\subsubsection{\textbf{Interest}}
\label{sec:daf-interest} 
It is similar to self-learning~\cite{Junxiao:SelfLearning:2017}. However, instead of dropping the Interest on a FIB mismatch, DAF sends it out as broadcast, similar to in-network RREQ broadcast in AODV. When a node receives an Interest with name \verb|/p/a|, it creates a PIT entry with tuple $<$$NONCE, ep, T(R_i)$$>$. The $ep$ is a unique identifier (e.g., MAC address) of the Interest forwarder (previous hop).

\subsubsection{\textbf{Data}}
\label{sec:daf-data-reply} 
A Data node replies upon receiving an Interest. On a broadcast Interest, it adds the producer prefix \verb|/p| with data. When a node receives a data, it performs a FIB lookup, measures the $RTT_i$ for $D_i$ on a match using Equation~\ref{eq:rtt}, and computes $SRTT(nh_p)$ and $RTTV(nh_p)$ using Equations~\ref{eq:srtt} and \ref{eq:rttv} respectively. Afterwards, it creates or updates a FIB entry for \verb|/p| with tuple,
\begin{equation*}
  <nh_p, hc, T(D_i), RTT_i, SRTT(nh_p), RTTV(nh_p)>.
\end{equation*} 
Here, $hc$ is the hop count between data and current node and is used to select the closet $nh$ to \verb|/p| for future Interest(s). A node then updates the CS \textit{before} PIT lookup for opportunistic caching. Afterward, it unicasts data if one $ep$ in the PIT entry for \verb|/p/a| or broadcasts for more than one and ignores otherwise. 

Note that with opportunistic caching on unsolicited data (no PIT entry), a FIB-next-hop entry may become stale by not being used for future unicast Interest. A periodic checker with a configurable timer (e.g., 1s) removes such stale entries using $T(D_i)$ as the starting time.

\section{Performance Analysis}
\label{sec:evaluation}

This section analyzes \textit{how} and \textit{why} the NDN architecture can perform better than IP in MANETs at the network layer and compare our DAF strategy against NDN flooding, self-learning, and IP-AODV schemes. We exclude any name-based forwarding outside the core NDN design like~\cite{Meisel:2010:LFBL}. We consider a client or consumer as a \textit{requester} and a server or data-node as a \textit{responder}. There are four performance metrics in our analysis,

a) \textit{Expected success rate (ESR)}: percentage of data packets retrieved by requester applications with respect to unique requests sent.

b) \textit{Average latency}: average round-trip time (RTT) at the requester applications. We measure RTT with respect to the first request packet when retransmission is enabled.

c) \textit{Network Tx-events}: the ratio of the total packets transmitted (Tx) by all nodes in a network to the total data packets retrieved by the requester applications; and, 

d) \textit{Average hop count}: average number of hops between requester-responder.

A forwarding strategy or routing protocol should yield high ESR with low latency, Tx-events, and average hop counts. Note that, for retrieving a single data packet at the application level, the number of data packet transmissions can be $> 1$ because of multi-hop traversal.
We simulated the four schemes using ndnSIM \cite{mastorakis2017ndnsim}. Each requester expects 500 different data packets of 512 bytes using a constant bit-rate (CBR) \textit{request-response} application at five requests per second. A requester may re-transmit (RTx=0 or max. 2) a request upon timeout. \cite{Meisel:2010:LFBL} explains the request-response application in IP-MANET. They argue that such an application is a proper fit for a MANET evaluation because sending data in a single direction with UDP CBR is indistinguishable from DDoS attacks. An IP request has twelve bytes, four for an increasing sequence number and eight for a timestamp.

A requester application starts randomly between 1.0 to 3.0 seconds and runs until it receives all 500 data packets or the final Interest times-out. RTS/CTS is disabled as a proper solution is not available for broadcast Interest and Data. In NDN, each node has a CS capacity of 0, 5, or 200 (default) packets. We also collect an average of 10 runs per simulation.


\subsection{Effect of node mobility or speed in AODV and DAF}

\begin{figure}[!t]
  \centering
  \subfloat[Expected success rate.]{
    \includegraphics[width=0.47\columnwidth]{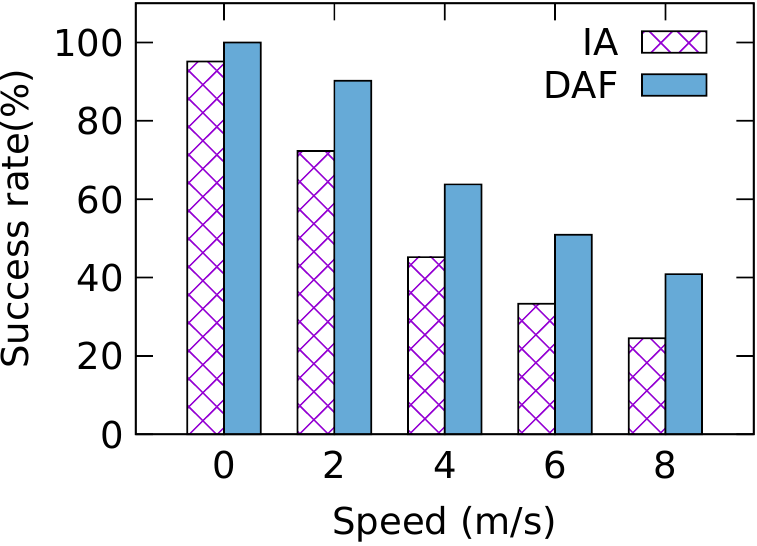}
    \label{fig:aodv-daf-esr-50}
  }
  \hfill
  \subfloat[Average latency.]{
    \includegraphics[width=0.47\columnwidth]{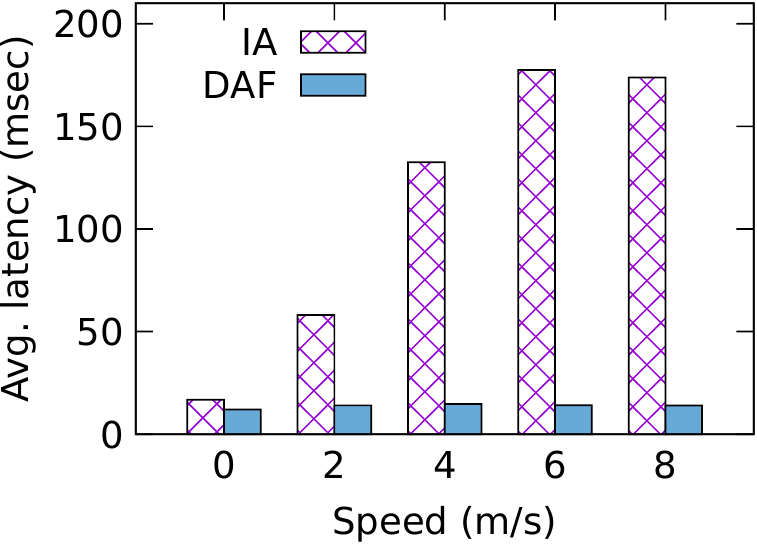}
    \label{fig:aodv-daf-latency-50}
  }
  \par 
  \subfloat[Tx events per data packet.]{
    \includegraphics[width=0.47\columnwidth]{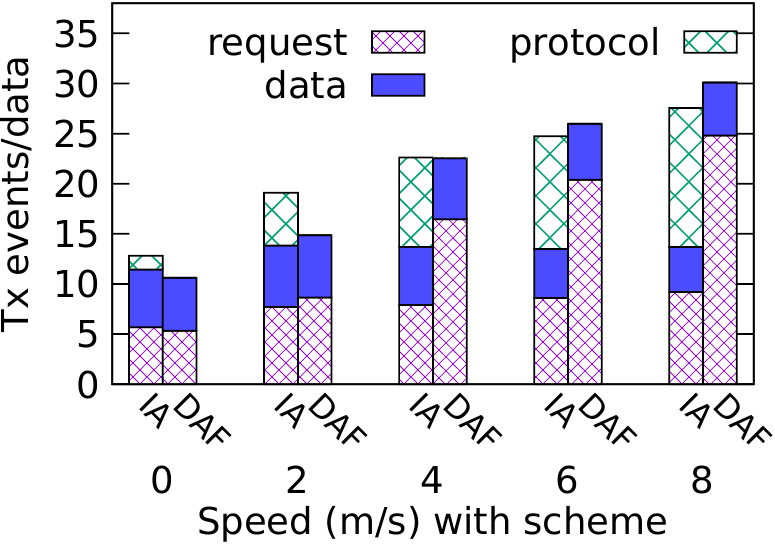}
    \label{fig:aodv-daf-tx-count-50}
  }
  \hfill
  \subfloat[Average hop count.]{
    \includegraphics[width=0.47\columnwidth]{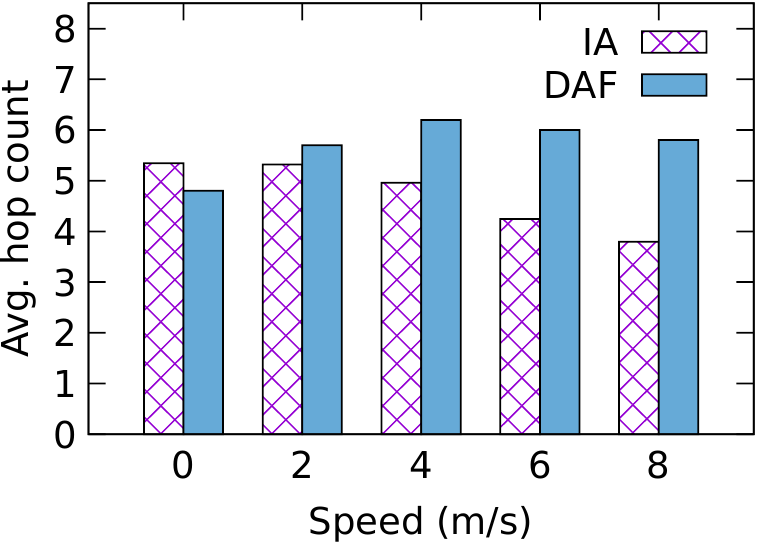}
    \label{fig:aodv-daf-hops-50}
  }
  \caption{Effect of mobility in IP-AODV (IA) and NDN-DAF (DAF) in 50 nodes network with 1-to-1 requester-responder communication at five requests per second. Retransmission is disabled and CS=200 packets per node for DAF.}
  \label{fig:aodv-daf-speed-50}
\end{figure}

We start by analyzing the effect of mobility in IP-AODV and NDN-DAF in a 50 node network in Fig.~\ref{fig:aodv-daf-speed-50} with a 1-to-1 communication (ten requesters and ten responders) and retransmissions disabled. In IP, this setup is apparent, with one requester address talking to a specific responder address. In NDN, each consumer requests the unique producer prefix's content, and each producer serves only one prefix content. The analysis also shows the lower-bound forwarding/routing performance of DAF against AODV as there is no caching or Interest aggregation effect for multicast. We also see the latency effect in a MANET and how lower latency helps DAF outperform AODV in terms of expected success rate (ESR). The increasing speed increases the possibility of link breakage, and thus the ESR gradually goes down for both AODV and DAF in Fig.~\ref{fig:aodv-daf-esr-50}. On average, DAF shows 38.1\% more ESR than AODV. It is because, on average, DAF has 75.5\% less latency in Fig.~\ref{fig:aodv-daf-latency-50} and 5.08\% fewer Tx-events per data retrieved in Fig.~\ref{fig:aodv-daf-tx-count-50} than AODV. 

At speed=0 m/s, AODV shows lower ESR than DAF because IP requires route set up \textit{before} data exchange, which incurs added delay (Fig.~\ref{fig:aodv-daf-latency-50}). Even the small number of protocol messages (Fig.~\ref{fig:aodv-daf-tx-count-50}) is good enough to induce delay and collisions/contentions during the initial route setup, causing packet loss at the beginning of application startup, and thus AODV's ESR is about 95\%. On the contrary, DAF can \textit{learn} a next-hop towards data only with Interest-Data packets without any protocol message exchange and achieves 100\% ESR.

AODV performance drops further under mobility because frequent link breakage increases protocol messages (Fig.~\ref{fig:aodv-daf-tx-count-50}). It also adds more delay from RREQ exponential back-off during route setup/maintenance, as shown in Fig.~\ref{fig:aodv-daf-latency-50} (maximum 25x than DAF). Higher latency in AODV also leads to nodes moving away before retrieving data. As a result, AODV retrieves most data when a requester is close to a responder, i.e., lower average hop-counts in Fig.~\ref{fig:aodv-daf-hops-50}). On the other hand, DAF maintains its lower latency with nodes' increasing moving speed as it learns path towards data with Interest packets. Thus, DAF can retrieve data over a longer path, contributing to its higher transmission events than AODV in Fig.~\ref{fig:aodv-daf-tx-count-50} under high mobility. We consider it a fair trade-off for obtaining higher ESR.

\begin{figure}[!ht]
  \centering
  \subfloat[Expected success rate.]{
    \includegraphics[width=0.47\columnwidth]{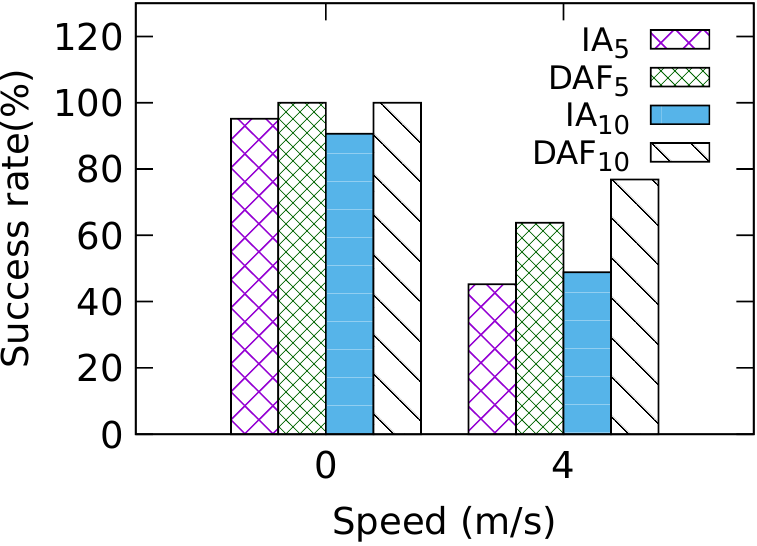}
    \label{fig:aodv-daf-esr-50-rate}
  }
  \hfill
  \subfloat[Average latency.]{
    \includegraphics[width=0.47\columnwidth]{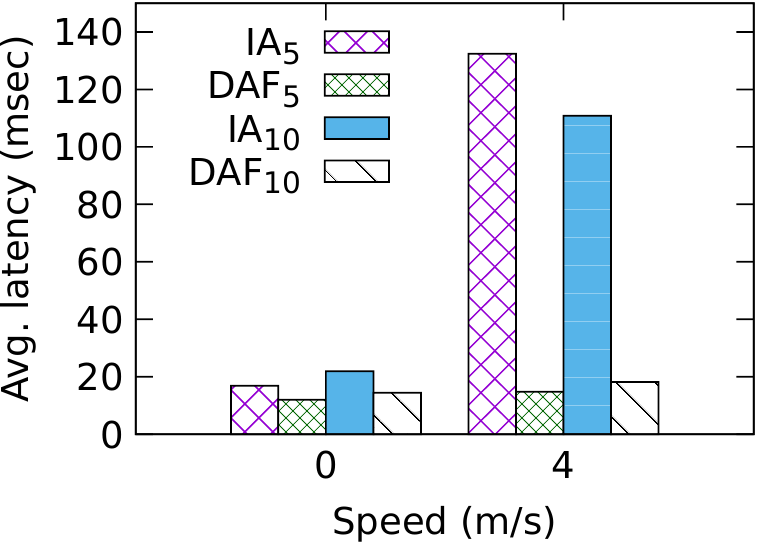}
    \label{fig:aodv-daf-latency-50-rate}
  }
  \par 
  \subfloat[Tx events per data packet.]{
    \includegraphics[width=0.47\columnwidth]{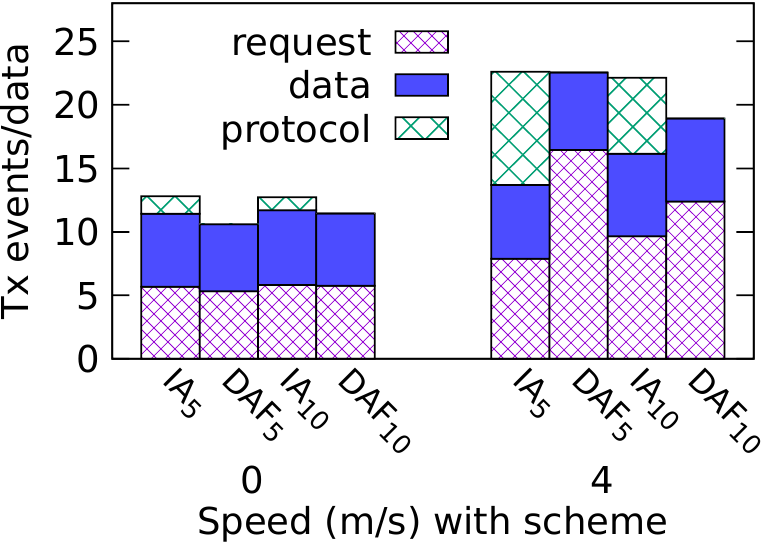}
    \label{fig:aodv-daf-tx-count-50-rate}
  }
  \hfill
  \subfloat[Average hop count.]{
    \includegraphics[width=0.47\columnwidth]{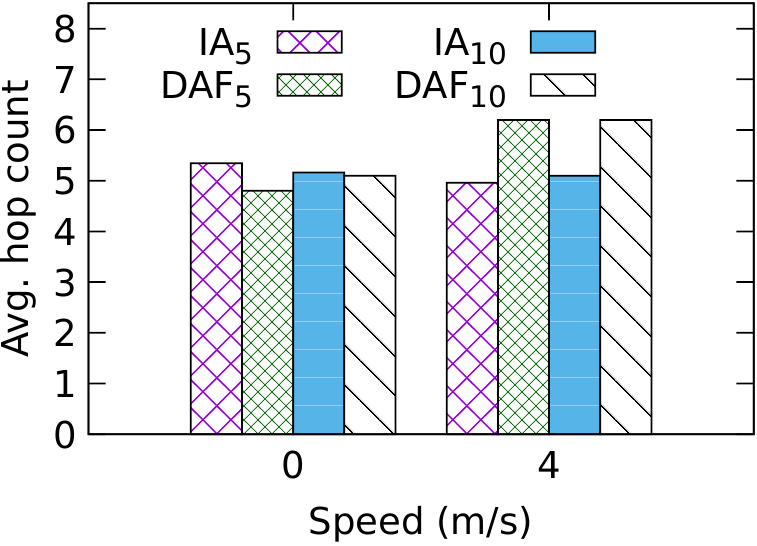}
    \label{fig:aodv-daf-hops-50-rate}
  }
  \caption{Validating mobility effect in IP-AODV (IA) and NDN-DAF (DAF) with different request rates in 50 nodes scenario with 1-to-1 requester-responder communication. Subscript shows requests per second. Retransmission is disabled and CS=200 packets per node for DAF.}
  \label{fig:aodv-daf-speed-50-rate}
\end{figure}

\textbf{AODV packet loss at startup:} We verify AODV's packet loss during the initial route setup by testing with a higher request sending rate in Fig.~\ref{fig:aodv-daf-speed-50-rate}. We consider the cases where the nodes are either stationary or moving at 4 m/s. Fig.~\ref{fig:aodv-daf-esr-50-rate} shows that when nodes are stationary, AODV's ESR drops to around 90\% at ten requests per second compared to 95\% at five requests per second. Higher request rates cause more packet loss due to higher collision and contention, causing further route setup delay. The added latency in Fig.~\ref{fig:aodv-daf-latency-50-rate} at ten requests/s compared to 5 requests/s proves our claim. 

Under mobility, however, AODV shows better ESR at a higher request rate. The protocol message exchange for \textit{route maintenance} goes down (Fig.~\ref{fig:aodv-daf-tx-count-50-rate}) as request-response exchange updates the routes' states more frequently. Moreover, the network sparsity goes higher under mobility in our simulation model, leading to less chance of collision. Thus, the latency goes down slightly (Fig.~\ref{fig:aodv-daf-latency-50-rate}), helping to quickly fetch data even from longer paths as seen in (Fig.~\ref{fig:aodv-daf-hops-50-rate}).

A similar effect is also visible in DAF, where a higher Interest rate fetches data more frequently, updating FIB entries in the process under fewer collision possibilities. However, the latency goes up by a small margin (Fig.~\ref{fig:aodv-daf-latency-50-rate}) in 10 Interests/s compared to 5 Interests/s as the opportunistic forwarding can take longer paths (Fig.~\ref{fig:aodv-daf-hops-50-rate}) slightly.

To summarize the design differences, AODV forwards data packets only over an active route between the source and destination IP addresses mandated by the IP design. On the other hand, NDN's topology-agnostic design allows DAF to opportunistically explore and repair the path between consumer and data node with in-network Interest broadcast. It also enables DAF to avoid any explicit route setup before data exchange, as seen in IP-AODV, offering a better performance trade-off in MANETs.

\subsection{NDN baseline for MANET}

Here, we analyze our DAF baseline forwarding performance in MANET and why we need it by comparing it against NDN-flooding and self-learning. 

\subsubsection{Effect of mobility}

\begin{figure}[!t]
  \centering
  \subfloat[Expected success rate.]{
    \includegraphics[width=0.47\columnwidth]{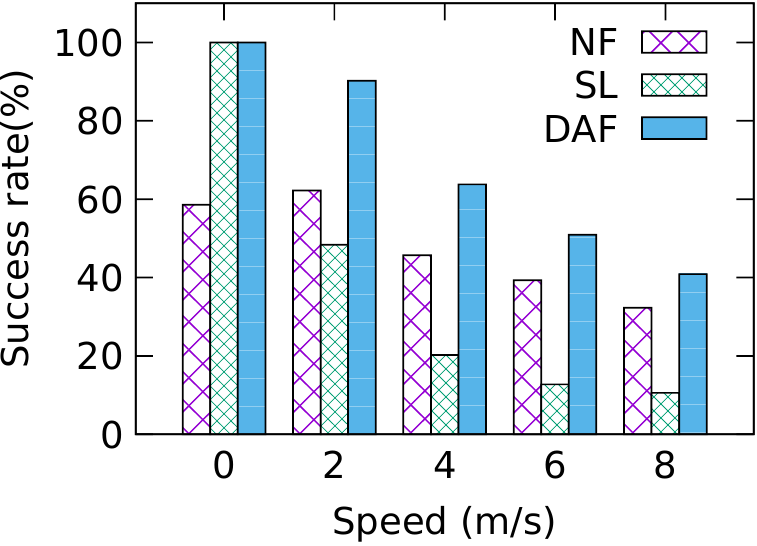}
    \label{fig:esr-50}
  }
  \hfill
  \subfloat[Tx events per data packet.]{
    \includegraphics[width=0.47\columnwidth]{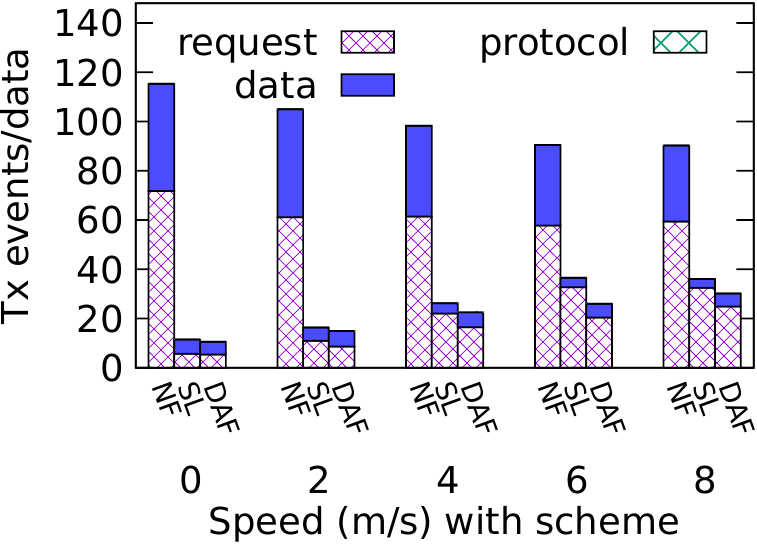}
    \label{fig:tx-count-50}
  }
  \caption{Effect of mobility in DAF, Flooding (NF), and Self-learning (SL) in 50 nodes scenario with 1-to-1 communication at five requests per second. Retransmission is disabled and CS=200 packets per node.}
  \label{fig:speed-50}
\end{figure}

Fig.~\ref{fig:speed-50} shows the limitations of flooding and self-learning and DAF's advantages in MANETs. The average latency and hop counts in 1-to-1 communication are very similar for all three schemes and thus not included in the figure. Fig.~\ref{fig:esr-50} shows that when nodes are stationary, the apparent effect of flooding leads to very high collision/contention resulting in only 59.8\% ESR. Fig.~\ref{fig:tx-count-50} verifies the high packet transmission. Self-learning at speed=0 m/s performs similar to DAF, given it also learns the next-hop to data only once. 

However, under mobility, self-learning's ESR degrades rapidly because only a consumer can broadcast Interest upon timeout at the application level. Flooding's ESR stays higher than self-learning because the network gets sparse under mobility in our simulation model, leading to less collision and better multi-path exploration. However, such performance comes at the cost of very high network transmission events.

DAF offers the right balance between flooding and self-learning with its RTT-based in-network unicast-broadcast interchange. It allows it to re-learn next-hops quickly and reduce overall network transmission cost. On average, DAF offers 42.28\% and 177.93\% more ESR than flooding and self-learning, respectively.

\begin{figure}[!t]
  \centering
  \subfloat[Expected success rate.]{
    \includegraphics[width=0.47\columnwidth]{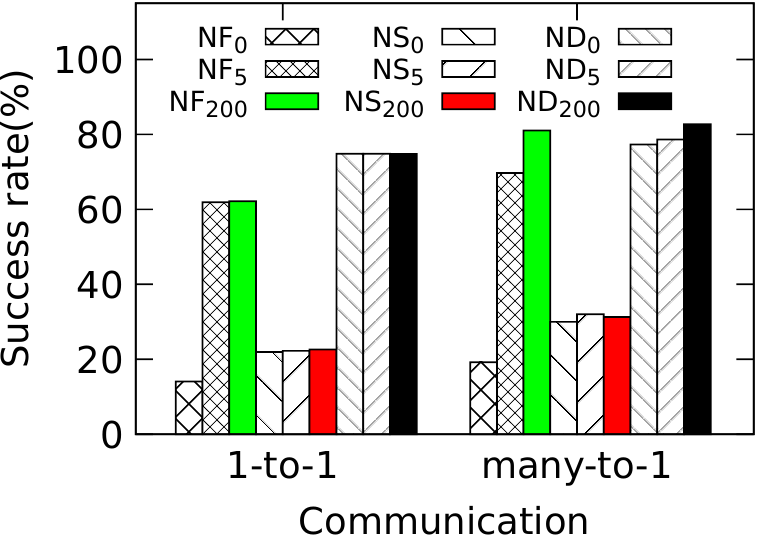}
    \label{fig:esr-cache}
  }
  \hfill
  \subfloat[Tx bytes per data packet.]{
    \includegraphics[width=0.47\columnwidth]{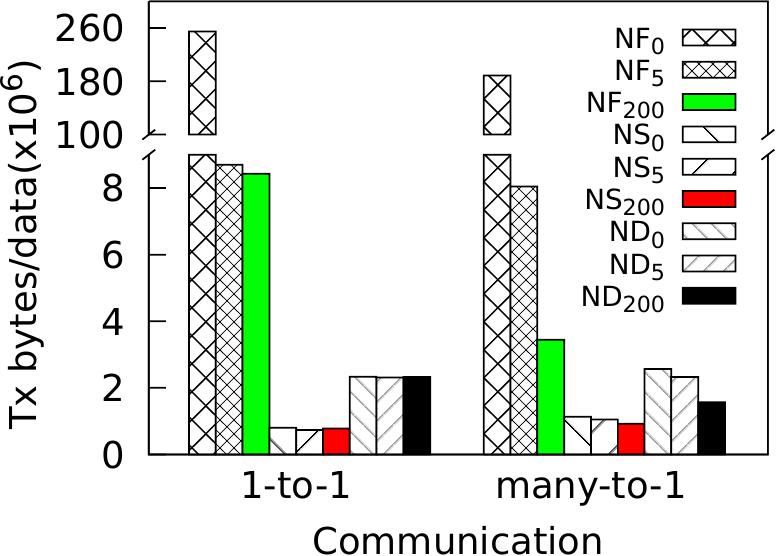}
    \label{fig:tx-bytes-cache}
  }
  \caption{Effect of content store (CS) in 1-to-1 and many-to-1 communication in 50 nodes scenario at speed=4 m/s with five Interests per second and application retransmissions (RTx) enabled (NDN-flooding: NF; NDN-self-learning: SL and NDN-DAF: DAF); Subscript shows CS capacity in packets.}
  \label{fig:cache-nodes}
\end{figure}

\subsubsection{Effect of Content Store (CS)}

To further strengthen our claim that DAF properly utilizes NDN's stateful forwarding plane, we analyze the effect of content-store (or CS) capacity in Fig.~\ref{fig:cache-nodes} at speed=10 m/s using a least-recently-used (LRU) eviction policy. Besides 1-to-1 communication, we also include the simulation scenario with many-to-1 communication where we have only two responders, one serving prefix \verb|/A|, and another serving \verb|/B|. Half of the requesters ask for data with prefix \verb|/A|, and the rest ask for data with prefix \verb|/B|. The goal is to observe the network behavior with caching and data multicast through Interest aggregation in NDN.

\textbf{1-to-1 communication}: Flooding at CS=0 has a 77.4\% less success rate than CS=200 (Fig.~\ref{fig:esr-cache}). Without CS, all Interests (including RTx) will always head for the actual producer leading to a broadcast storm. Fig.~\ref{fig:tx-bytes-cache} shows very high Tx-bytes/data in flooding, causing higher collision/contention. However, caching significantly lowers flooding Tx-bytes and improves ESR. Note that at CS=5, flooding has lower Tx-bytes than CS=200. It is because large CS (late eviction) creates a higher number of disjoint paths than small CS (early eviction). Caching shows a minimal effect on self-learning and DAF as all the data packets are unicast in a 1-to-1 setup. Thus, only the nodes along a data-requester path can cache, and a retransmitted Interest hardly hits a CS under mobility.

\textbf{Many-to-1 communication}: Here, flooding ESR in all three CS capacities is higher than their respective 1-to-1 bars (Fig.~\ref{fig:esr-cache}). The effect of Interest aggregation and multicast is evident here, which significantly lowers the Tx-bytes (Fig.~\ref{fig:tx-bytes-cache}). Self-learning again does not gain much success rate with caching as most Interests are lost or dropped. Thus Tx-bytes per data is also the lowest as a consumer retrieves most data packets when it is closer to the data node. Our DAF strategy appropriately reacts to link failure with in-network Interest broadcast on a FIB mismatch. The opportunistic caching from data broadcast with NDN multicast helps DAF improve ESR with increasing cache capacity, mitigating flooding's broadcast storms and self-learning's higher Interest loss. 

The analysis above validates our proposed NDN baseline forwarding in a mobile network. In the following sections and subsections, we will only consider DAF and AODV for further comparative analysis between NDN and IP in MANETs.

\subsection{Effect of network size or number of nodes}
\label{subsec:effect-network-size}

\begin{figure}[!t]
  \centering
  \subfloat[Expected success rate.]{
    \includegraphics[width=0.47\columnwidth]{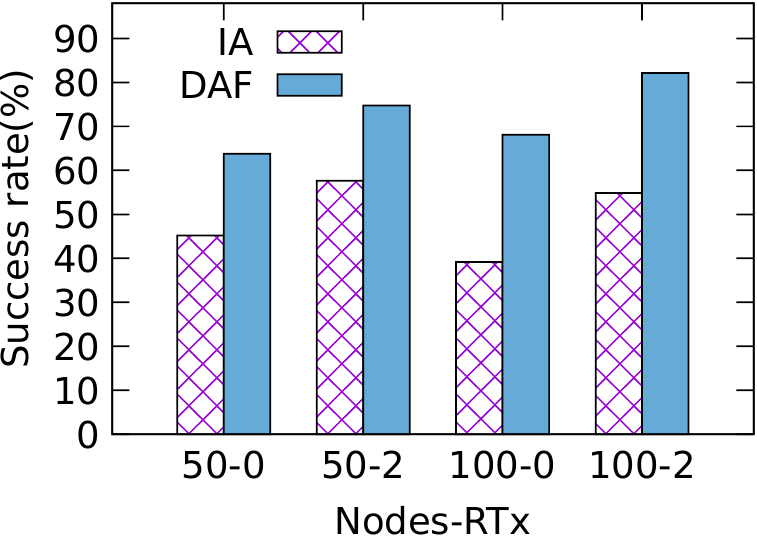}
    \label{fig:esr-var}
  }
  \hfill
  \subfloat[Average latency.]{
    \includegraphics[width=0.47\columnwidth]{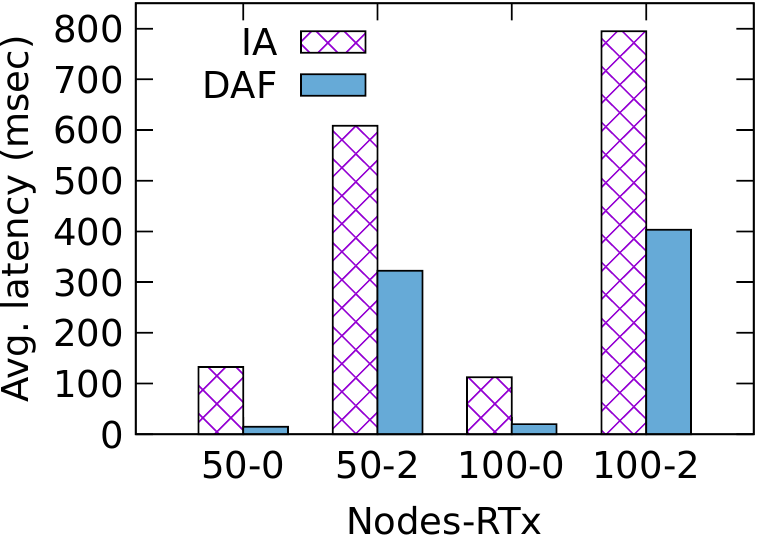}
    \label{fig:latency-var}
  }
  \par
  \subfloat[Tx events per data packet.]{
    \includegraphics[width=0.47\columnwidth]{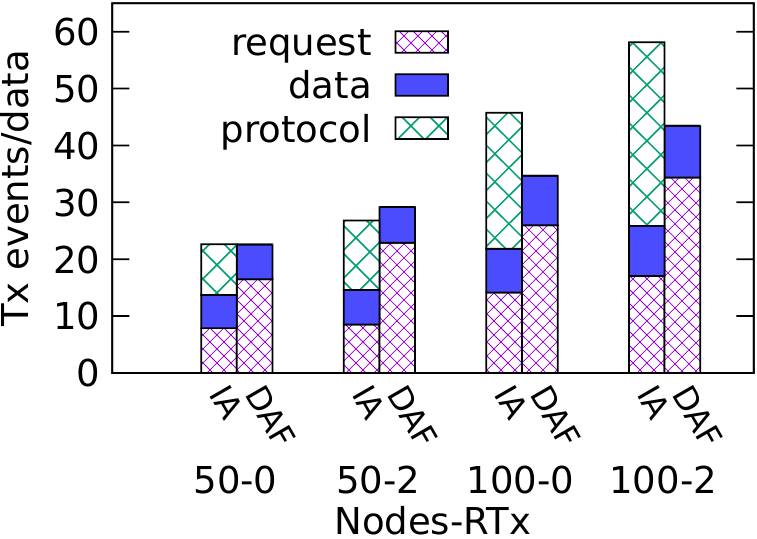}
    \label{fig:tx-count-var}
  }
  \hfill
  \subfloat[Average hop count.]{
    \includegraphics[width=0.47\columnwidth]{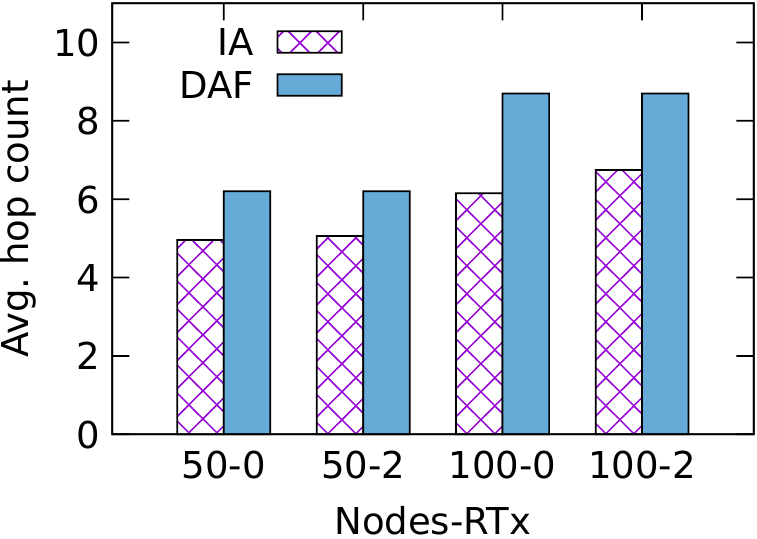}
    \label{fig:hops-var}
  }
  \caption{Effect of network size on AODV (IA) and DAF with 1-to-1 communication at speed=4 m/s with five requests per second and application retransmission (RTx) enabled and disabled; CS=200 packets per node in DAF.}
  \label{fig:var-nodes}
\end{figure}

Fig.~\ref{fig:var-nodes} shows the effect of network size (or nodes) and provides the worst-case comparison between the DAF and AODV under mobility. A cache hit is impossible without RTx unless a broadcast Interest finds cached data. A tuple X-Y in the x-axis shows X = the number of nodes and Y = RTx.

At similar RTx, the latency (Fig.~\ref{fig:latency-var}) and Tx-events (Fig.~\ref{fig:tx-count-var} of both schemes go up when the network size increases. A requester is further away from a responder on a more extensive network (Fig.~\ref{fig:hops-var}). At RTx=2, both DAF and AODV get better ESR at the cost of extra Tx. The latency also increases as our application-level timeout is 2s, and we measure RTT with respect to the first request when RTx is enabled.

A more extensive network at the same RTx worsens AODV's ESR in Fig.~\ref{fig:esr-var}. Protocol messages also increase, introducing longer delays. However, DAF's Interest-Data-only feedback-reaction mechanism reduces the latency by an average of 69.7\% compared to AODV. It also has fewer request packets than AODV's protocol and requests messages combined (Fig.~\ref{fig:tx-count-var}). With lower latency and Tx-events, DAF achieves an average of 45.46\% more ESR (Fig.~\ref{fig:esr-var}) than AODV. Note that DAF shows an average of 27.03\% higher hop count than AODV (Fig.~\ref{fig:hops-var}) as the lower latency enables it to retrieve data quicker under mobility. It leads to an average of 6.54\% more transmitted-bytes per data than AODV (not shown in plots). However, we consider it a fair trade-off, given DAF reduces Tx-events and improves the success rate. Such performance comes from better utilizing the underlying NDN design.

\subsection{Effect of many-to-many communication}
\label{subsec:effect-m2m}

\begin{figure}[!t]
  \centering
  \subfloat[Expected success rate.]{
    \includegraphics[width=0.47\columnwidth]{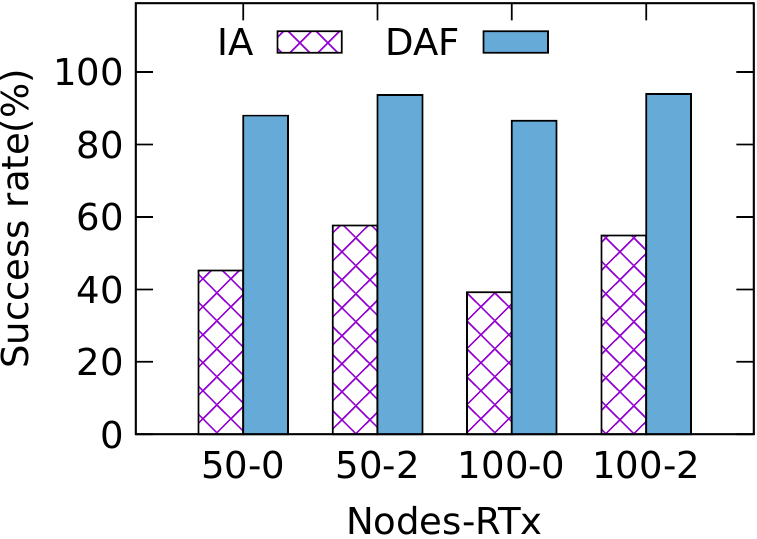}
    \label{fig:esr-m2m}
  }
  \hfill
  \subfloat[Average latency.]{
    \includegraphics[width=0.47\columnwidth]{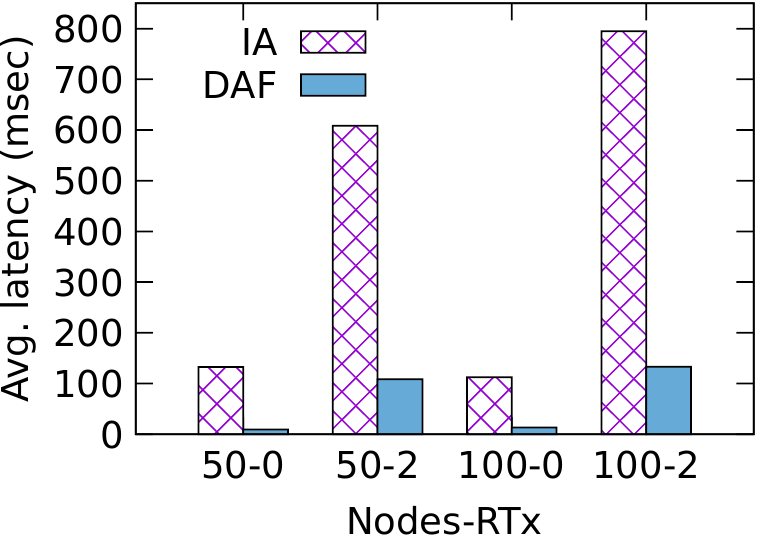}
    \label{fig:latency-m2m}
  }
  \par
  \subfloat[Tx events per data packet.]{
    \includegraphics[width=0.47\columnwidth]{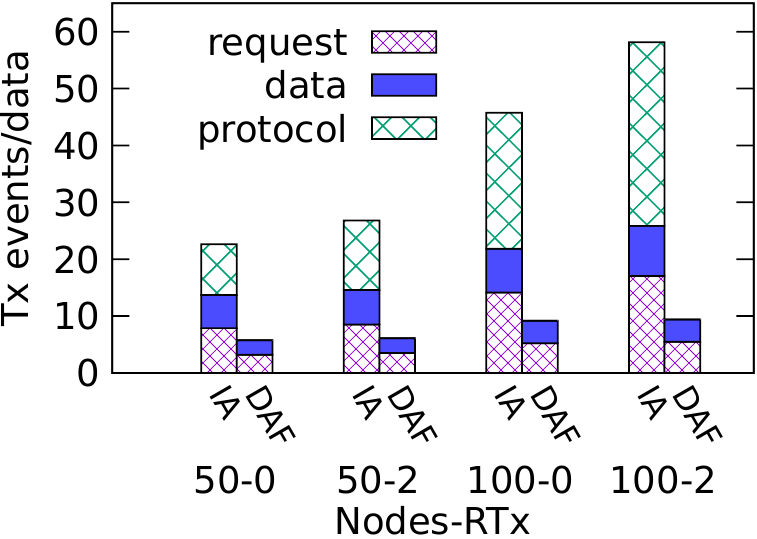}
    \label{fig:tx-count-m2m}
  }
  \hfill
  \subfloat[Average hop count.]{
    \includegraphics[width=0.47\columnwidth]{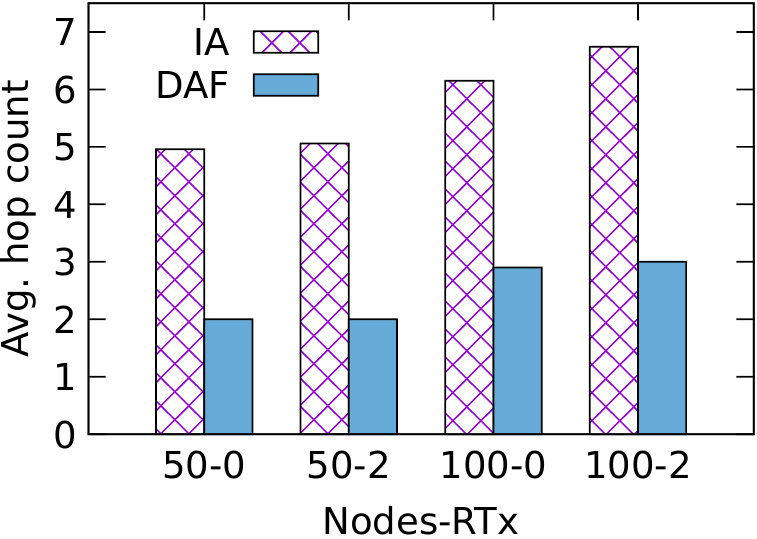}
    \label{fig:hops-m2m}
  }
  \caption{Effect of many-to-many requester-responder communication on AODV (IA) and DAF at speed=4 m/s with five requests per second and application retransmission (RTx) enabled and disabled; CS=200 packets per node in DAF.}
  \label{fig:m2m-nodes}
\end{figure}

Next, we consider that 5 out of 10 responders serve data with the same prefix (e.g., \verb|/A|), while the other half serve another prefix (e.g., \verb|/B|). Similarly, half of the requesters ask for data with prefix \verb|/A|, and the rest for prefix \verb|/B|. Even though various responders can serve the same data in IP, one requester address can always communicate with only one responder IP address. It provides a best-case example for NDN forwarding against IP routing in MANET. The results are in Fig.~\ref{fig:m2m-nodes}.

We can see that DAF outperforms AODV with higher ESR under all setups in Fig.~\ref{fig:esr-m2m}. The direct effect of NDN's location-agnostic design, built-in caching and multicast is now visible. Given the network’s responsibility to retrieve data rather than reach a specific IP address, DAF’s feedback-reaction mechanism learns the opportunistic shortest path to an eligible data node, including cached data. When a path to such a data node breaks (or a cache miss occurs), especially under mobility, DAF can quickly react to it and switch to another data node without any protocol message using its in-network RTT measurement. As a result, DAF achieves an average of 88.09\% less latency compared to AODV. Opportunistic caching with data broadcast through NDN multicast helps DAF lower the hop count by 57.58\% in Fig.~\ref{fig:hops-m2m}, which then helps lower the Tx-events by an average of 80.14\% compared to AODV in Fig.~\ref{fig:tx-count-m2m}. AODV cannot use the multicast and caching features without a highly-modified protocol. It provides the same results as 1-to-1 communication because the underlying address-based IP architecture lacks any application-level knowledge. 

Combining the effects mentioned above, DAF achieves, on average, 82.98\% more success rate (Fig.~\ref{fig:esr-m2m}) than AODV. Although DAF shows lower average hops than AODV, the possible sub-optimal longer path formation still results in higher hops than flooding and self-learning (not shown in plots). Note that flooding can achieve an average of 85\% more ESR than AODV in the many-to-many scenario (not shown in plots) as opportunistic data caching can lead hop counts to be zero for many data packets. However, the broadcast storm still leads to very high Tx-events in the network, which DAF mitigates with its in-network unicast-broadcast interchanging.

\subsection{Effect of many-to-1 communication}
\label{subsec:effect-many-to-1}

\begin{figure}[!t]
  \centering
  \subfloat[Expected success rate.]{
    \includegraphics[width=0.47\columnwidth]{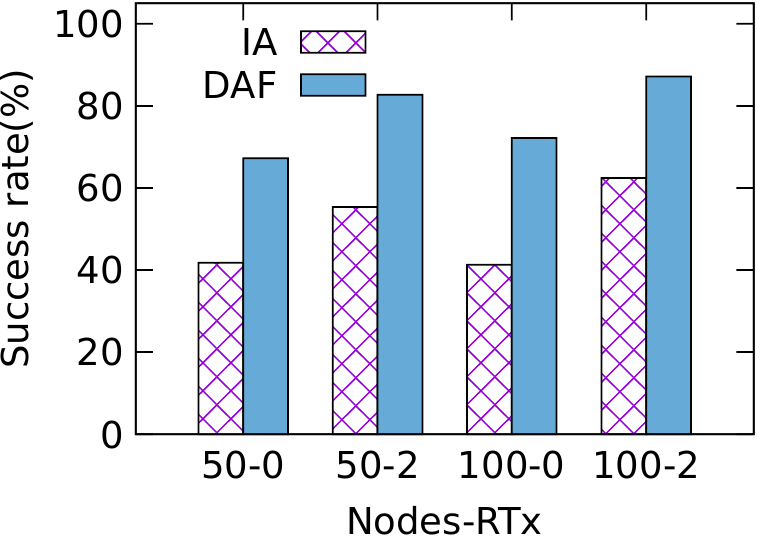}
    \label{fig:esr-2p}
  }
  \hfill
  \subfloat[Average latency.]{
    \includegraphics[width=0.47\columnwidth]{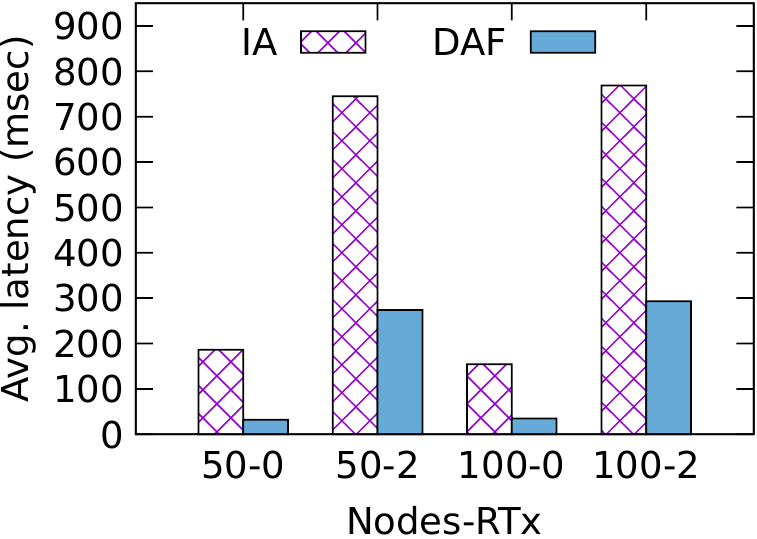}
    \label{fig:latency-2p}
  }
  \par
  \subfloat[Tx events per data packet.]{
    \includegraphics[width=0.47\columnwidth]{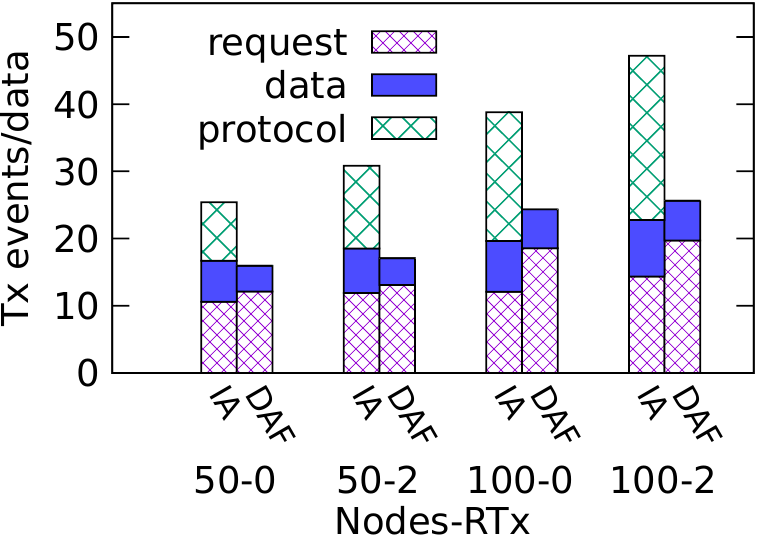}
    \label{fig:tx-count-2p}
  }
  \hfill
  \subfloat[Average hop count.]{
    \includegraphics[width=0.47\columnwidth]{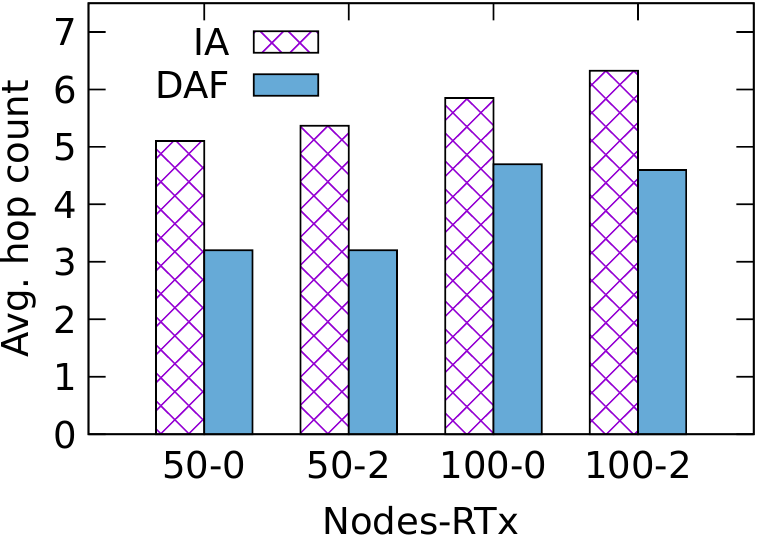}
    \label{fig:hops-2p}
  }
  \caption{Effect of many-to-1 requester-responder communication on AODV (IA) and DAF at speed=4 m/s with five requests per second and application retransmission (RTx) enabled and disabled; CS=200 packets per node in DAF.}
  \label{fig:2p-nodes}
\end{figure}

Here, we have only two responders, one serving prefix \verb|/A|, and the other \verb|/B|. Ten requesters ask for content in a similar way to Section.~\ref{subsec:effect-m2m}. The goal is to observe the network behavior with a higher traffic load at and near the actual data responder or \textit{``hot zones."} Fig.~\ref{fig:2p-nodes} shows the simulation results.

DAF shows an average of 50.35\% more success rate than AODV in Fig.~\ref{fig:esr-2p}. Data broadcast from PIT aggregation reduces dedicated per downstream Tx and helps with opportunistic caching in overhearing neighbors. Thus, DAF reduces the traffic load at and near a producer with NDN's built-in multicast when multiple consumers ask for the same content. As a result, DAF's TX-events per data, latency, and hop count goes down by an average of 41.66\% (Fig.~\ref{fig:tx-count-2p}), 61.85\% (Fig.~\ref{fig:latency-2p}), and 27.31\% (Fig.~\ref{fig:hops-2p}), respectively, compared to AODV. DAF also has 29.98\% lower Tx-bytes (not shown). Compared to DAF's underlying NDN architecture, the underlying IP model of AODV requires per requester dedicated flow for the same responder even though all requesters ask for the same content. Thus we can say that NDN offers a better baseline forwarding than IP in a MANET.

%
\section{Challenges and Future Work}
\label{sec:challenges-future-works}

To maintain a baseline forwarder design, we do not consider a collision avoidance (random delay timer or back-off) mechanism in DAF. We believe that a better trade-off between success rate and latency is possible with proper modifications under mobility. Our baseline may also require further optimization, e.g., using directionality in vehicular networks, better reliability in disaster, or mission-critical networks. Actual device implementation of DAF, along with testing, including multiple channels and NICs, should provide valuable insights into NDN's real-world performance. 

We also conducted experiments with in-network retransmissions in DAF in a 50 node 1-to-1 communication network to improve the success rate. However, it yielded feeble ESR gain ($<$ 1\%) over baseline DAF and increased packet transmission per data packet by over 25\%. This overhead occurs when multiple nodes get a timeout on their FIB-next-hops at the same time. Thus, each node ends up sending out in-network retransmission as broadcast. This effect produces a skyrocketing number of Interests, duplicated data packets from the producer(s) and cached node(s), leading to increased collision and contention. Thus we did not consider this approach in our analysis.

\section{Conclusion}
\label{sec:conclusion}
This work provides a baseline comparison between NDN and IP in the MANET environment. IP-based protocols need to maintain working paths between the sending and receiving mobile hosts, which is inherently challenging in MANET, causing significant latency and transmission overhead. By moving away from the point-to-point model to a data retrieval model, NDN no longer requires path maintenance between two specific endpoints; it allows retrieving requested data from anywhere in the network, thus enables caching, multicast, and stateful forwarding. Our DAF strategy is a baseline NDN-based MANET forwarding mechanism. It reduces the network latency and, in most cases, the transmission overhead, thus improving application packet retrieval over IP-AODV. While our work demonstrates NDN architecture's advantages for MANET, further improvements and optimizations to the baseline design are possible.


\bibliographystyle{IEEEtran}  
\bibliography{refs}

\end{document}